\documentclass[11pt]{article}
\usepackage{epsfig}
\usepackage{axodraw}

\begin{document}
\normalsize
\title{\Large  Testing the Structure of the Scalar Meson $K_0^*(1430)$ in
  $\tau\to K_0^*(1430)\nu$ Decay}
\author{Mao-Zhi Yang\\
{\small CCAST (World Laboratory), P.O. Box 8730, Beijing 100080, China}\\
{\small and }\\
{\small Institute of High Energy Physics, Chinese
Academy of
Sciences,}\\
 {\small P.O. Box 918(4), Beijing 100049, P.R. China}\footnote{Mailing address}
}
\date{\empty}
\maketitle

%\begin{picture}(0,0)
%       \put(340,230){BIHEP-TH-2005-}
%       \put(340,210){\bf }
%\end{picture}

\begin{abstract}
The decay constant of $K_0^*(1430)$ is the key quantity to
determine the production rate of $K_0^*(1430)$ in $\tau$ decays.
By assuming $K_0^*(1430)$ is the lowest scalar bound state of
$s\bar{q}$, the decay constant can be calculated reliably in QCD
sum rule. Then the decay branching ratio of $\tau\to
K_0^*(1430)\nu$ is predicted to be about $(7.9\pm 3.1)\times
10^{-5}$. If this branching ratio can be measured by experiment,
it should be helpful to make clear the structure of $K_0^*(1430)$.
\end{abstract}

\hspace{1cm} \small{PACS numbers: 13.35.Dx}

\hspace{2cm}

The structure of light scalar mesons is still a common problem for
physics of light hadrons. There are too many light scalar mesons
to be accommodated into one $SU(3)$ nonet. There are: 1) $\kappa$
and $K_0^*(1430)$ with strange number $|S|=1$ and isospin $I=1/2$,
2) $\sigma$, $f_0(980)$, $f_0(1370)$, $f_0(1500)$ and $f_0(1710)$
with both strange number and isospin 0, and 3) $a_0(980)$,
$a_0(1450)$ with isospin $I=1$ \cite{pdg2004,ishida}. It is most
possible that these scalar mesons make up at least two nonets. One
is below 1GeV, the other is above 1GeV. Which of them belong to
the $q\bar{q}$ scalar nonet is still an unsolved problem
\cite{close1}. Up to now, there is still no general agreement for
the structure of light scalar mesons. They can be understood as
four-quark states, meson-meson molecular states
\cite{JAFFE,ISGUR}, or ordinary quark-antiquark bound states. From
the point of view of quark model and QCD, there must be scalar
mesons composed of quark and antiquark.

$\kappa$ and $K_0^*(1430)$ have the same strange number and
isospin, but different masses. Therefore, they must belong to
different nonets. Whether they are composed of quark-antiquark, if
they are, where the lowest scalar bound state of quark-antiquark
exists are interesting questions. To answer these questions, a
large amount of experiments and theoretical analysis need to be
done to analyze the production and decay properties of the scalar
mesons. In this note, we calculate the production rate of
$K_0^*(1430)$ in the heavy lepton $\tau$ decay process $\tau\to
K_0^*(1430)\nu$. There is only one hadron evolved in this decay
process, the interaction in the leptonic vertex can be calculated
with high precision, therefore the decay branching ratio depends
on the decay constant of $K_0^*(1430)$ directly. The decay
constant is determined by the meson's structure. Therefore, if the
branching ratio can be measured by experiment, comparing the
theoretical prediction with experimental data will indicate
information on the structure of the scalar meson.

We analyzed the mass of the lowest scalar bound state of two
quarks $s\bar{q}$ with QCD sum rule, where $q=u,d$, denotes light
quarks \cite{massk}. We find that it is impossible to obtain the
mass of $\kappa$ in QCD sum rule for $s\bar{q}$ bound state.
However, QCD sum rule can give stable prediction of the mass of
$K_0^*(1430)$. Therefore, we assume that $K_0^*(1430)$ is the
lowest scalar bound state of $s\bar{q}$, and $\kappa$ is
irrelevant to $s\bar{q}$ scalar channel. With this assumption, the
decay constant of $K_0^*(1430)$ is calculated to be
\begin{equation}
f_{K_0^*}=427\pm 85 \mathrm{MeV} \label{con}
\end{equation}
which is defined by the matrix element
\begin{equation}
\langle 0|\bar{q}s|K_0^*(1430)\rangle =m_{K_0^*}f_{K_0^*}
\end{equation}
where $m_{K_0^*}$ is the mass of $K_0^*(1430)$.

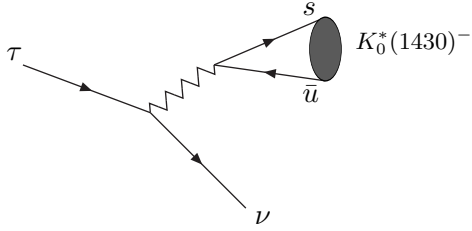
\begin{figure}[h]
\begin{center}
\begin{picture}(120,90) (84,-60)
\SetWidth{0.5} \SetColor{Black} \ArrowLine(84,6)(132,-12)
\ArrowLine(132,-12)(168,-48) \ZigZag(132,-12)(156,6){3}{4}
\ArrowLine(156,6)(198,24) \ArrowLine(198,0)(156,6)
\GOval(198,12)(12,6)(0){0.3528}

\put(78,8){$\tau$} \put(172,-54){$\nu$}
\put(190,25){$s$}\put(190,-8){$\bar{u}$}
\put(210,12){\footnotesize{$K_0^{*}(1430)^-$} }
\end{picture}
\caption{Feynman diagram for $\tau\to K_0^*(1430)\nu$}
\end{center}
\end{figure}
The diagram for $\tau\to K_0^*(1430)\nu$ is shown in Figure 1.
According to the diagram, we can write the decay amplitude
directly
\begin{equation}
A=\frac{ig^2}{8}V^*_{us}\langle K_0^*(1430)|\bar{s}\gamma_\mu
(1-\gamma_5) u|0\rangle \frac{1}{p_1^2-m_W^2+i\epsilon}
\bar{\nu}\gamma^\mu(1-\gamma_5)\tau \label{A}
\end{equation}
where $g$ is the weak coupling constant, $V_{us}$ is the CKM
matrix element, and $p_1$ the momentum of $K_0^*(1430)$. Let us
denote the momenta of $\tau$ lepton and neutrino as $p_{\tau}$ and
$p_2$, respectively.

From the parity of $K_0^*(1430)$ and the axial vector current, we
know that the contribution of axial vector current to the matrix
element must be zero
\begin{equation}
\langle 0|\bar{u}\gamma_\mu\gamma_5 s|K_0^*(1430)\rangle =0
\label{axial}
\end{equation}
therefore, only the contribution of vector current should be
considered.  From the Lorentz property of the matrix $\langle
0|\bar{u}\gamma_\mu s|K_0^*(1430)\rangle $, we can write
\begin{equation}
\langle 0|\bar{u}\gamma_\mu s|K_0^*(1430)\rangle
=f^V_{K_0^*}p_{1\mu}\label{vect}
\end{equation}
where $f^V_{K_0^*}$ is a new constant defined for the scalar
meson.

Using Dirac equation, we can get the following operator equation
\begin{equation}
\partial^\mu(\bar{u}\gamma_\mu s)=i(m_u-m_s)\bar{u}s
\end{equation}
then we get the following relation
\begin{eqnarray}
i(m_u-m_s)\langle 0|\bar{u} s|K_0^*(1430)\rangle &=\langle
0|\partial^\mu(\bar{u}\gamma_\mu s)|K_0^*(1430)\rangle \nonumber\\
&=-ip_1^\mu \langle 0|\bar{u}\gamma_\mu s|K_0^*(1430)\rangle
\end{eqnarray}
From the above relation, and using the definition of the decay
constant $f_{K_0*}$ and $f_{K_0*}^V$, we get
\begin{equation}
f_{K_0*}^V=\frac{m_s-m_u}{m_{K_0^*}}f_{K_0*}\label{fv}
\end{equation}
Then considering eqs.(\ref{axial}, \ref{vect},
\ref{fv}), from
eq.(\ref{A}) the result of the decay amplitude squared is
\begin{equation}
\sum_{\lambda_1\lambda_2}|A|^2=|\frac{G_F}{\sqrt{2}}V_{us}^*
 \frac{m_s-m_u}{m_{K_0^*}}f_{K_0^*}|^22m_{\tau}(m_{\tau}^2-m_{K_0^*}^2)
 \end{equation}
where $G_F$ is Fermi constant, $\lambda_1$ and $\lambda_2$ are the
polarization freedom of $\tau$ lepton and neutrino, respectively.

Finally we obtain the decay width of $\tau\to K_0^*(1430)\nu$
\begin{equation}
\Gamma=\frac{1}{4\pi}|\frac{G_F}{\sqrt{2}}V_{us}^*
 \frac{m_s-m_u}{m_{K_0^*}}f_{K_0^*}|^2
 \frac{(m_{\tau}^2-m_{K_0^*}^2)^2}{2m_{\tau}}
\end{equation}
The branching ratio of this decay mode is defined as
\begin{equation}
Br(\tau\to K_0^*(1430)\nu)=\Gamma/\Gamma_{total}
\end{equation}
The total decay width of $\tau$ lepton is related to its mean life
time as
\begin{equation}
\Gamma_{total}=\hbar/\tau
\end{equation}

In the numerical calculation, we take
\begin{eqnarray}
\tau &=&2.9\times 10^{-13}\mathrm{s},
~~~~V_{us}=0.22\nonumber\\
m_{K_0^*}&=&1.412\mathrm{GeV}~~~~~m_{\tau}=1.777\mathrm{GeV}\nonumber\\
m_s &=&0.14\mathrm{GeV},~~~~~~~~m_q\sim 0 \nonumber
\end{eqnarray}
With the input parameters taken above, and the decay constant
calculated in QCD sum rule (eq.(\ref{con})), the result of the
branching ratio is
\begin{equation}
Br(\tau\to K_0^*(1430)\nu)=(7.9\pm 3.1)\times 10^{-5}
\end{equation}
where the error bar is mainly contributed by the uncertainty of
the decay constant, and the effects of the other uncertainties are
negligible.

The above prediction is several times smaller than the present
experimental upper limit \cite{pdg2004}
\begin{equation}
Br(\tau\to K_0^*(1430)\nu)^{Exp.}<5\times 10^{-4}
\end{equation}
It is most possible that this decay branching ratio can be
measured in $e^+ e^-$ collision experiments with high luminosity,
such as CLEO-c, the incoming BESIII, or the asymmetic B-factories.
If this branching ratio can be measured in the future, the
comparison of the experimental data with theoretical prediction,
which is obtained by assuming that $K_0^*(1430)$ is the lowest
scalar bound state of quark-antiquark pair $s\bar{q}$, will give
some information about the component of this scalar meson. Then we
may know where the $q\bar{q}$ $SU(3)$ nonet exists in the
spectroscopy of light scalar mesons.

\vspace{1cm}

{\bf Acknowledgements} This work is supported in part by the
National Science Foundation of China under contract No.10205017,
and by the Grant of BEPC National Laboratory.

\end{document}